\input{aipcheck.tex}

\def\selectedoptions{final}

\documentclass[
   \selectedoptions
  ]
  {aipproc}

\layoutstyle{6x9}

\SetInternalRegister\hbadness{8000} 

\newcommand\doingARLO[2][]{%
  \ifx\mmref\undefined #1\else #2\fi
}

\begin{document}

\title 
      {Hadron-Hadron Scattering in the Nonrelativistic Quark Model}


\author{T.Barnes}{
address={Physics Division, Oak Ridge National Laboratory, 
Oak Ridge, TN 37831-6373, USA \\ 
Department of Physics and Astronomy, University of Tennessee  
Knoxville, TN 37996-1501, USA },
email={barnes@bethe.phy.ornl.gov}
}

\begin{abstract}
In this HADRON2001 
contribution we summarize the status of our quark-model calculations of
hadron-hadron scattering amplitudes in annihilation-free channels.
The predictions are in reasonably good agreement with
experimentally known S-wave meson-meson and meson-baryon phase shifts, and
there are very recent indications that S-wave $\pi \omega$ scattering
(extracted from FSIs in $b_1$ decay) may also be similar to our predictions.
Finally, novel applications of this formalism to the dissociation 
cross sections of charmonia on light hadrons 
(relevant for QGP studies at RHIC) are discussed.
\end{abstract}

\date{\today}

\maketitle

\section{Hadron-hadron scattering}

\subsection{Introduction}

Early models of strong interhadron forces were constructed by analogy with 
QED Feynman diagrams, and
in the important NN problem it was assumed that these forces
were dominated by $t$-channel meson exchange.
At large distances, one-pion-exchange 
can indeed be confirmed in NN high partial waves. 
However at short distances 
one must have serious reservations about this type of
model, since exchange of a ca. 1~GeV meson in $t$-channel implies a range of
about 0.2~fm. 
Since this is much smaller than the extent of a typical hadron, the
assumption of $t$-channel meson exchange appears rather 
dubious (see Maltman and Isgur 
\cite{Maltman/Isgur:1984}
for a discussion). The success of meson-exchange models
may simply be due to the many parameters available for fitting, and less
trivially because it may prove difficult to distinguish short-distance 
QCD processes such as quark interchange 
from meson exchange, as these involve the same flavor flow. 

Since QCD is a theory of quarks and gluons, and short-ranged
scattering probes hadronic
wavefunctions, it may be possible to describe
hadron-hadron scattering in terms of explicit quark model wavefunctions and interquark
forces taken from QCD and hadron spectroscopy. This approach has a long history
in the NN problem,
and has been applied to many hadronic reactions with much success
and the occasional interesting failure. The technique most often used is the
resonating-group method, although other variational or
nonperturbative methods have also been applied. 
In this contribution we discuss our
results from much simpler Born-order calculations of scattering amplitudes
in the quark model, 
which are more straightfoward to evaluate and are also in reasonable agreement
with experimental S-wave scattering amplitudes.

\subsection{Quark Born Diagrams}

The usual quark model interaction has $\lambda\cdot\lambda$ color dependence,
and in consequence 
a single interaction between quarks in different hadrons transforms
the incident hadronic clusters
from color singlets to color octets. 
Although this makes direct (no quark exchange)
scattering zero at Born order, this modified state does have overlap with
final color singlet hadrons, provided that we allow quark interchange.
We refer to the resulting diagrams as ``quark Born diagrams".
The four quark Born diagrams one finds for the scattering of two $q\bar q$
mesons through this mechanism are
shown in Fig.1.
(We label these diagrams
according to type; if the
interacting constituents scatter into the same final hadron this is a
``capture" diagram, and if not it is a ``transfer" diagram.)

Each diagram has an associated spatial 
overlap integral, which is weighted by color, spin
and flavor multiplicative matrix elements. 
Detailed evaluation of these diagrams was discussed in Refs.\cite{BS92,S92},
and the ``Feynman rules" for the hadron-hadron T-matrices in our current
notation are given in Ref.\cite{PsV}.
With Gaussian wavefunctions one may evaluate the T-matrices
and phase shifts in closed form. 
As an example, the I=2 $\pi\pi$ S-wave phase shifts
from standard quark model interactions (Ref.\cite{PsV}) 
are give in Eq.\eqref{pipiSwave}.

\begin{figure}
  \resizebox{16pc}{!}{\includegraphics[height=.3\textheight]{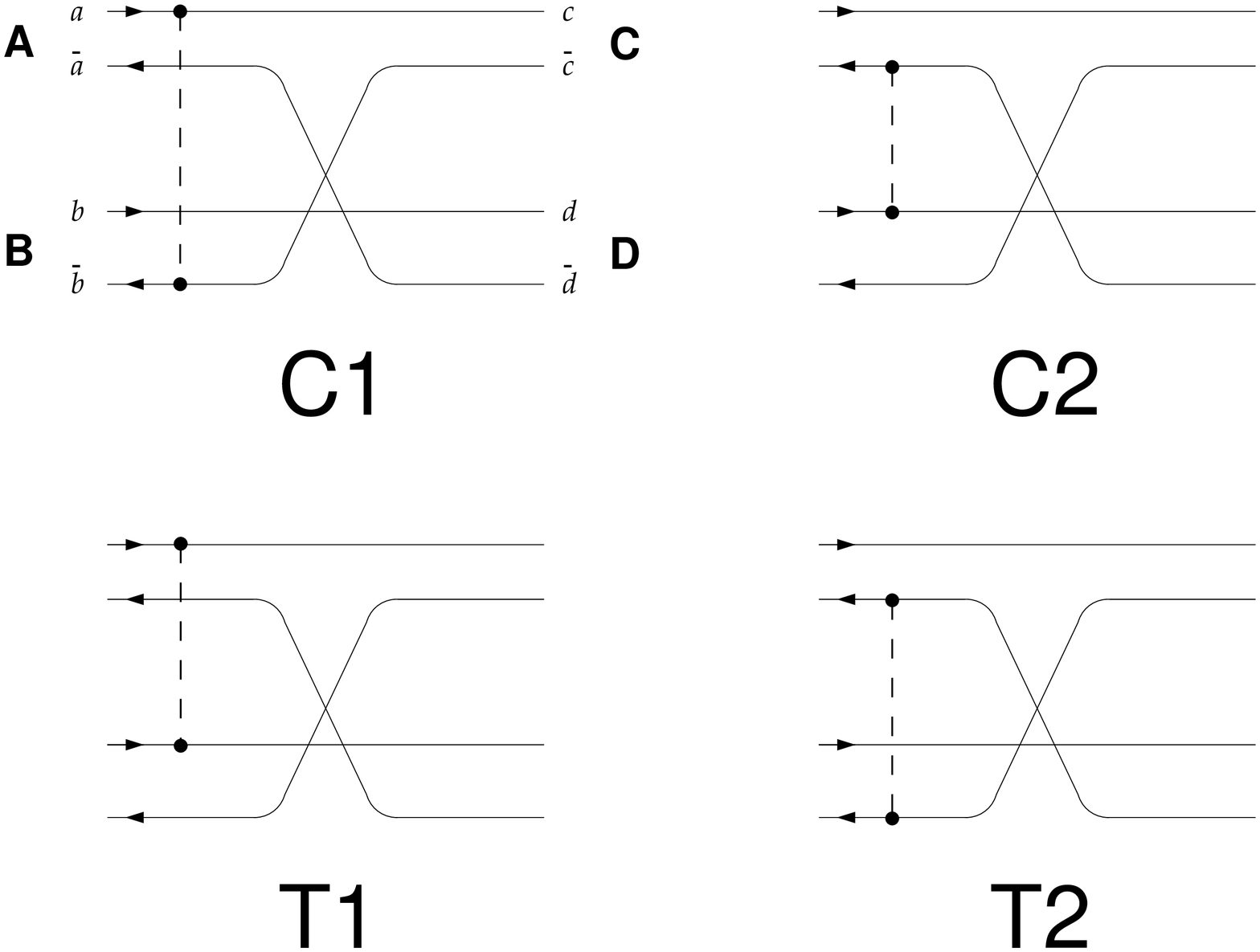}}
  \caption{The four quark Born diagrams for meson-meson scattering.}
\end{figure}

\begin{equation}
\delta^{I=2\ \pi\pi}_0
=
\cases{
kE_{\pi}
{\alpha_s\over m^2}\;
\bigg(
-{1\over 3^2}\;
{1\over x}
\Big(
1-e^{-2x}
\Big)
-{2^4\over 3^{7/2}}\,  e^{-4x/3}
\bigg)
&{\rm OGE $S\cdot S$}
\cr
kE_{\pi}
{\alpha_s \over \beta^2}\;
\bigg(
-{2\over 3^2}
{1\over x}
\Big(
{\rm f}_{1,{1\over 2}}(-2x)
-e^{-2x}
\Big)
-{2^3 \over 3^{5/2}}\;
{\rm f}_{1,{3\over 2}}(-2x/3)\, e^{-4x/3}
\bigg)
\,
\;
&{\rm color Cou.} \cr
kE_{\pi}
{b\over \beta^4}\;
\bigg(
{1\over 3^2}
{1\over x}
\Big(
{\rm f}_{2,{1\over 2}}(-2x)
-e^{-2x}
\Big)
+
{1\over 3^{1/2}}\;
{\rm f}_{2,{3\over 2}}(-2x/3)\, e^{-4x/3}
\bigg)\
\,
\;
&{\rm lin. conft.}
}
\label{pipiSwave}
\end{equation}
where
$x = {\vec A\,}^2 / 4\beta^2$, 
${\rm f}_{a,c}(x)$
is an abbreviation for
the confluent hypergeometric function
${}_1{\rm F}_1(a;c;x)$,
$|\vec A\, |$ is the pion momentum in the c.m. frame (we assume
a relativistic dispersion relation), $\beta=0.4$~GeV is
the standard quark model $q\bar q$ wavefunction width parameter, and a 
conventional quark model parameter set of $\alpha_s=0.6$, $m_q=0.33$~GeV and 
$b=0.18$~GeV$^2$ is used to give the curves in Fig.2.
The total Born-order S-wave phase shift is the sum of these
three contributions. We have confirmed that these results are
quite similar to the variational results of Weinstein
and Isgur \cite{Weinstein/Isgur:1990}, 
who used essentially the same interactions but included 
contributions beyond Born order. 

Application of this approach to the scattering of other hadron pairs is
straightforward, one need only enumerate the complete set
of Born-order scattering diagrams, and evaluate these given a set of
external hadron wavefunctions. We have applied this method to S-wave 
scattering of a wide range of annihilation-free channels, specifically
I=3/2
K$\pi$ \cite{Kpi},
I=0,1 KN \cite{KN},
I=0,1 BB \cite{BB} (compared to LGT data),
and
the NN repulsive cores \cite{NN},
with generally
reasonable results.
These references consider many additional cases for which we do not have data
at present.

More sensitive tests of the hadron scattering mechanism are possible if we
consider higher partial waves. Here there is evidence of very 
interesting physics, for example in the large NN spin-orbit force 
(referred to by Isgur as the ``Holy Grail"
of quark-model scattering calculations) and the similarly large
KN spin-orbit force. The KN spin-orbit force is apparently not well explained 
as elastic scattering with quark model forces \cite{KN1}. (N.Black,
unpublished, finds very similar results to this reference.) This discrepancy
with experiment may be due to the large inelasticities known 
experimentally to be present in KN scattering.

\begin{figure}
  \resizebox{18pc}{!}{\includegraphics[height=.3\textheight]{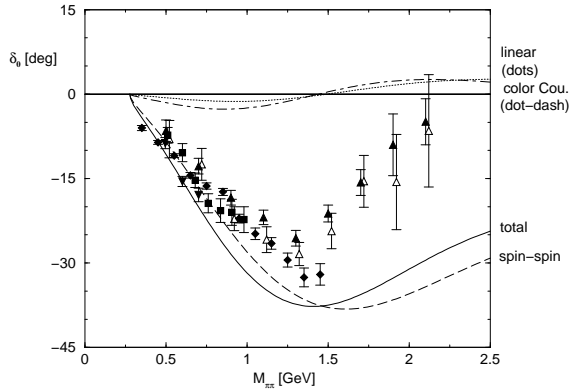}}
  \caption{I=2 $\pi\pi$ experimental S-wave phase shifts versus
Eq.\eqref{pipiSwave}, from Ref.\cite{PsV}.}
\end{figure}

We recently considered light 
vector-pseudoscalar meson scattering
as a model spin-orbit problem for the quark Born diagram 
formalism, and derived the complete set of phase shifts in all
partial waves given Gaussian wavefunctions and standard quark model forces
\cite{PsV}. We found that quark-model spin-orbit effects 
can indeed be quite large,
for example in P-wave I=2 $\rho\pi$ elastic scattering we
found a phase shift splitting $\delta(^3{\rm P}_2) - \delta(^3{\rm P}_0)$ 
that peaked at
about $40^\circ $. Although one might suppose vector-pseudoscalar scattering 
to be 
experimentally inaccessible, it actually can be 
measured as a final state
interaction in multiamplitude decays. In $b_1\to\omega\pi$ in particular
the S and D amplitudes have FSI phases of 
$e^{i\delta_S}$
and 
$e^{i\delta_D}$, so the 
S-D cross term in the $\omega\pi$ angular distribution is suppressed by
$\cos(\delta_S - \delta_D)$ relative to the 
$|S|^2$
and 
$|D|^2$ terms. Our prediction is that
\begin{equation}
\delta_S(\omega\pi) - \delta_D(\omega\pi) = -14^\circ
\end{equation}
at the $b_1$ mass. The E852 Collaboration 
has used this unusual FSI technique to extract this relative phase,
and finds a consistent result of
$\delta_S(\omega\pi) - \delta_D(\omega\pi) 
\approx -19^\circ (4^\circ )(8^\circ )$. 
($\delta_D - \delta_S =  0.327 \pm 0.061\pm 0.143$ radians is the 
tentative, as yet unpublished 
E852 result; private communication from M.Nozar.)

\subsection{A New Application: Charmonium Dissociation at RHIC}

Recently a novel class of hadronic reactions has attracted 
the attention of physicists searching for evidence of quark gluon plasma
formation in heavy ion collisions.
One signature proposed as an indicator 
of QGP formation is a suppression
of the production rate of charmonium bound states 
such as the $J/\psi$, since the
QGP is expected to screen the linear potential that 
would normally encourage a $c\bar c$
pair produced in the collision to remain bound \cite{Matsui:1986}. 

If the charmonia that are formed in the
collisions can penetrate the cloud of "comoving" 
light hadrons also produced in the
collision, they can be detected through characteristic decays such as 
$J/\psi \to \ell^+\ell^-$, and this will be a ``clean" experiment.
Alternatively, if inelastic charmonium + light hadron cross sections
into open-charm final states
are sufficiently large, this more conventional $c\bar c$ dissociation 
process may imitate the expected QGP signal and 
will complicate the interpretation of the experiment.
Of course these charmonium + 
light hadron dissociation cross sections
are not at all well known at low energies, and estimates of the scale of these 
cross sections assuming different theoretical scattering mechanisms
cover many orders of magnitude. Here we may have an exciting 
opportunity to establish
the preferred hadron-hadron scattering mechanism in a new regime of QCD.

\begin{figure}
  \resizebox{22pc}{!}{\includegraphics[height=.3\textheight]{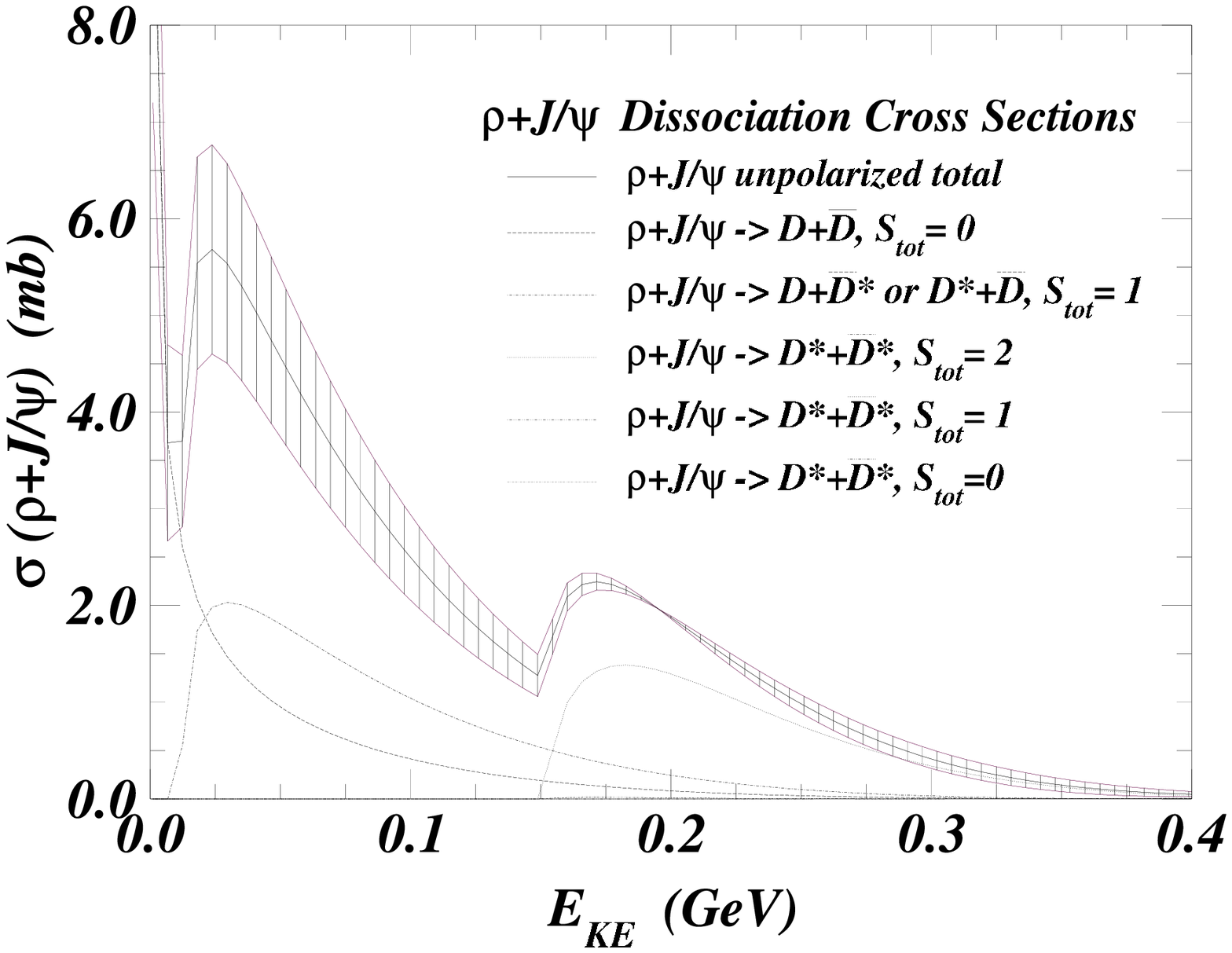}}
  \caption{The $\rho + J/\psi$ dissociation cross section predicted by 
quark Born
diagrams \cite{WSB:2001}.} 
\end{figure}

We have carried out a series of calculations of these 
charmonium + light meson dissociation cross sections
in the constituent interchange model, 
using the approach described above.
We use a standard Coulomb + linear + smeared hyperfine Hamiltonian 
to determine wavefunctions and scattering amplitudes, which are evaluated using 
numerical techniques. Our results are that these cross sections at leading order
are dissociation to open charm rather than elastic scattering (which is obvious
from the flavor flow in Fig.1), and the low-energy cross sections are typically
ca. 1~mb in scale \cite{WSB:2001,WSB:2000}. 
(Earlier work by Martins, Blaschke and Quack \cite{Martins:1995} using the same
scattering formalism found somewhat larger cross sections, due to 
their assumption of a color-independent confining interaction.)
Interestingly, 
$\rho + J/\psi$
cross sections are much larger than
$\pi + J/\psi$, in part because $\rho + J/\psi\to {\rm D} \bar {\rm D}$ 
is exothermic and hence 
diverges as we approach threshold. 
This process may lead to considerable suppression
of the initial $J/\psi$ population, 
and will require careful consideration in
applications of this charmonium-suppression idea to QGP searches.

In future we plan to extend these calculations 
to a wide range of initial and final
states, so that total cross sections (summed over all 
accessible final states) 
can be evauated, and more complicated reactions
such as N$+J/\psi$ 
inelastic scattering can be treated. Many other fascinating 
questions, 
such as the possibility that charmed mesons might bind to nucleons and 
nuclei, 
can also be considered through the application of this model of low-energy
interhadron forces.

\begin{theacknowledgments}
It is a great pleasure to thank the organisers of HADRON2001 for providing 
the opportunity to present these results and discuss them with my colleagues in
hadron physics. 
This work was supported in part by
the DOE Division of Nuclear Physics,
at ORNL,
managed by UT-Battelle, LLC, for the US Department of Energy
under Contract No. DE-AC05-00OR22725, and by the US National Science
Foundation under Grant No. INT-0004089.
\end{theacknowledgments}


\doingARLO[\bibliographystyle{aipproc}]
          {\ifthenelse{\equal{\AIPcitestyleselect}{num}}
             {\bibliographystyle{arlonum}}
             {\bibliographystyle{arlobib}}
          }
\bibliography{scat}

\hyphenation{Post-Script Sprin-ger}
\begin{thebibliography}{14}
\expandafter\ifx\csname natexlab\endcsname\relax\def\natexlab#1{#1}\fi
\providecommand{\enquote}[1]{``#1''}
\expandafter\ifx\csname url\endcsname\relax
  \def\url#1{\texttt{#1}}\fi
\expandafter\ifx\csname urlprefix\endcsname\relax\def\urlprefix{URL }\fi

\bibitem[K.Maltman and N.Isgur(1984)]{Maltman/Isgur:1984}
K.Maltman, and N.Isgur, \emph{Phys. Rev.}, \textbf{D29}, 952 (1984).

\bibitem[T.Barnes and E.S.Swanson(1992)]{BS92}
T.Barnes, and E.S.Swanson, \emph{Phys. Rev.}, \textbf{D46}, 131 (1992).

\bibitem[E.S.Swanson(1992)]{S92}
E.S.Swanson, \emph{Ann. Phys. (NY)}, \textbf{220}, 73 (1992).

\bibitem[T.Barnes et~al.(2001)]{PsV}
T.Barnes, N.Black, and E.S.Swanson, \emph{Phys. Rev.}, \textbf{C43}, 025204
  (2001).

\bibitem[J.Weinstein and N.Isgur(1990)]{Weinstein/Isgur:1990}
J.Weinstein, and N.Isgur, \emph{Phys. Rev.}, \textbf{D41}, 2236 (1990).

\bibitem[T.Barnes et~al.(1992)]{Kpi}
T.Barnes, E.S.Swanson, and J.Weinstein, \emph{Phys. Rev.}, \textbf{D46}, 4868
  (1992).

\bibitem[T.Barnes and E.S.Swanson(1994)]{KN}
T.Barnes, and E.S.Swanson, \emph{Phys. Rev.}, \textbf{C49}, 1166 (1994).

\bibitem[T.Barnes et~al.(1999)]{BB}
T.Barnes, N.Black, D.J.Dean, and E.S.Swanson, \emph{Phys. Rev.}, \textbf{C60},
  045202 (1999).

\bibitem[T.Barnes et~al.(1993)]{NN}
T.Barnes, S.Capstick, M.D.Kovarik, and E.S.Swanson, \emph{Phys. Rev.},
  \textbf{C48}, 539 (1993).

\bibitem[S.Lemaire et~al.(2002)]{KN1}
S.Lemaire, J.Labarsouque, and B.Silvestre-Brac, \emph{Nucl. Phys.}, \textbf{A}
  (2002).

\bibitem[T.Matsui and H.Satz(1986)]{Matsui:1986}
T.Matsui, and H.Satz, \emph{Phys. Lett.}, \textbf{B178}, 416 (1986).

\bibitem[C.Y.Wong et~al.(2001)]{WSB:2001}
C.Y.Wong, E.S.Swanson, and T.Barnes, {Heavy Quarkonium Dissociation Cross
  Sections in Relativistic Heavy-Ion Collisions}, Tech. Rep. nucl-th/0106067
  (2001).

\bibitem[C.Y.Wong et~al.(2000)]{WSB:2000}
C.Y.Wong, E.S.Swanson, and T.Barnes, \emph{Phys. Rev.}, \textbf{C62}, 045201
  (2000).

\bibitem[K.Martins et~al.(1995)]{Martins:1995}
K.Martins, D.Blaschke, and E.Quack, \emph{Phys. Rev.}, \textbf{C51}, 2723
  (1995).

\end{thebibliography}

\end{document}